\documentclass[
 amsmath,amssymb,
 reprint,
 superscriptaddress,aps,
 nofootinbib,
 prd,
 preprintnumbers,
]{revtex4-2}
\usepackage[english]{babel}
\usepackage[table]{xcolor}
\usepackage{graphicx}
\usepackage{amsmath}
\usepackage{amssymb}
\usepackage{bm}
\usepackage{dsfont}
\usepackage{bbm}
\usepackage{hyperref}
\usepackage{float}
\usepackage{color}
\usepackage[utf8]{inputenc}
\usepackage{amsmath,amsfonts}
\usepackage{booktabs}
\usepackage{longtable}
\usepackage{array}
\usepackage{makecell}
\usepackage{xspace}
\usepackage{multirow}
\usepackage{lineno}
\nolinenumbers

\newcommand{\SupplementalMaterial}{Supplemental Material~\cite{arxivSupplementalMaterial}}

\makeatletter
\newcommand{\supplementaltableofcontents}{%
  \begingroup
  \setcounter{tocdepth}{2}
  \@starttoc{stc}%
  \endgroup
}

\newcommand{\addsuppsection}[1]{%
  \addcontentsline{stc}{section}{\protect\numberline{\thesection}#1}
}
\newcommand{\addsuppsubsection}[1]{%
  \addcontentsline{stc}{subsection}{\protect\numberline{\thesubsection}#1}
}
\newcommand{\addsuppsubsubsection}[1]{%
  \addcontentsline{stc}{subsubsection}{\protect\numberline{\thesubsubsection}#1}
}
\makeatother

\begin{document}

\preprint{MITP-26-009}

\author{Aleks Smolkovi\v c}%
\affiliation{%
Jo\v zef Stefan Institute, Jamova 39, 1000 Ljubljana, Slovenia
}%
\author{Peter Stangl}
\affiliation{Institute of Physics, Johannes Gutenberg University Mainz, Staudingerweg 7, 55128 Mainz, Germany}

\title{Differentiable Multi-scale Effective Field Theory Likelihoods\\ for Beyond the Standard Model Phenomenology}

\begin{abstract}
Probing heavy new physics beyond the Standard Model (SM) increasingly relies on global effective field theory (EFT) likelihoods. We introduce differentiable, multi-scale EFT likelihoods that combine renormalization-group evolution, matching, observable predictions, and experimental constraints in a single differentiable framework. This enables modern gradient-based frequentist and Bayesian inference in large parameter spaces. We demonstrate these capabilities in two 374-parameter SMEFT analyses, making basis-independent, fully multi-scale global EFT analyses feasible in practice.
\end{abstract}


\maketitle

\section{Introduction}\label{sec:introduction}

The absence of direct observations of heavy new particles beyond the Standard Model has placed particle physics in an era of indirect searches for new physics, driven by increasing luminosity at the LHC and by numerous precision measurements at low energies. New physics with masses substantially above directly accessible energy scales can be systematically described by effective field theories (EFTs), which provide a controlled expansion in inverse powers of the new-physics scale and allow the inclusion of quantum corrections and the resummation of large logarithms through the renormalization group (RG). As a consequence, modern phenomenological analyses aiming to interpret experimental data in the context of heavy new physics are naturally formulated in terms of EFTs. The most widely used EFTs for this purpose are the Standard Model Effective Field Theory (SMEFT)~\cite{Buchmuller:1985jz,Grzadkowski:2010es,Brivio:2017vri,Isidori:2023pyp,Aebischer:2025qhh} above the electroweak scale and the Weak Effective Theory (WET)~\cite{Buchalla:1995vs, Jenkins:2017jig} below.

EFTs play a dual role in beyond-the-Standard-Model~(BSM) phenomenology. On the one hand, they provide a model-independent parameterization of new-physics effects at the energy scales relevant for the measured processes; on the other hand, explicit ultraviolet~(UV) models are connected to EFTs through matching calculations, allowing EFT analyses to serve as an essential intermediate step in phenomenological studies of full UV models.
In this context, an EFT likelihood quantifies the agreement between new-physics scenarios and experimental data and forms the central object for both model-independent EFT analyses and phenomenological studies of explicit UV models matched to the EFT.

A global approach to EFT phenomenology is essential. Individual Wilson coefficients typically contribute to many observables, while each observable depends on several Wilson coefficients. Moreover, RG evolution induces mixing among Wilson coefficients over the wide range of energy scales spanned by experimental data. The intricate relationships between effects in measurements at different scales and across experiments, together with the opportunity to combine them consistently in a single framework, require an approach that is global both in the space of Wilson coefficients and in the set of observables.

Despite significant progress~\cite{Aebischer:2018iyb,Hartland:2019bjb,Ellis:2020unq,Ethier:2021bye,Allwicher:2023shc,terHoeve:2023pvs,Celada:2024mcf,Greljo:2022jac,Buras:2024mnq,Bartocci:2024fmm, terHoeve:2025gey, deBlas:2025xhe,Mantani:2025bqu,Hirsch:2025qya, Greljo:2023adz, Greljo:2023bdy, Greljo:2023bab, Brod:2022bww,Kosnik:2025srw,Kley:2021yhn,Fajfer:2023gie,Brivio:2022hrb,Elmer:2023wtr, Grunwald:2023nli}, state-of-the-art EFT analyses remain limited in practice. Current approaches typically rely on low-dimensional analyses, strong flavor assumptions, or computationally expensive RG evolution and sampling or profiling methods. This leaves large regions of the parameter space unexplored, introduces spurious basis dependence, excludes many available observables, and hinders systematic connections to UV-complete models. These limitations are not conceptual but computational: they arise from the difficulty of performing high-dimensional, multi-scale analyses with existing likelihood implementations and tools.

In this work, we introduce a new paradigm for EFT phenomenology based on fully differentiable, multi-scale EFT likelihoods. By formulating the complete likelihood, including RG evolution, matching, and observable predictions, as a differentiable function of the Wilson coefficients using automatic differentiation~\cite{Baydin:2015tfa,jax2018github}, we enable the use of modern gradient-based inference techniques. This allows analytic evaluation of likelihood gradients and curvature, efficient optimization, and gradient-based Monte Carlo sampling in high-dimensional parameter spaces. We demonstrate how differentiable EFT likelihoods overcome key bottlenecks of traditional approaches and open new directions for EFT-based BSM phenomenology.

\section{Differentiable Multi-Scale EFT Likelihoods}\label{sec:formalism}

In existing global EFT analyses, practical limitations and computationally expensive sampling or profiling techniques typically necessitate a reduction of the parameter space, either by restricting attention to a small set of specific Wilson coefficients or by imposing flavor assumptions. The former leads to spurious basis dependence and the omission of potentially important correlations. Flavor assumptions introduce strong model dependence into an otherwise model-independent framework and, due to their instability under SMEFT RG evolution~\cite{Jenkins:2013zja,Jenkins:2013wua,Alonso:2013hga,Machado:2022ozb}, result in a spurious scale dependence. In both cases, large regions of the full EFT parameter space remain unexplored, and many observables that are sensitive to the omitted operators cannot be consistently included in the likelihood. Any such basis-, scale-, or model-dependent assumptions obstruct a systematic connection to generic UV models.

We address these limitations by constructing EFT likelihoods directly as functions of the full set of Wilson coefficients, without imposing flavor assumptions or restricting the parameter space a priori. Renormalization group evolution~\cite{Jenkins:2013zja,Jenkins:2013wua,Alonso:2013hga,Aebischer:2017gaw,Jenkins:2017dyc}, matching~\cite{Aebischer:2015fzz,Jenkins:2017jig}, and theoretical predictions for observables are incorporated consistently within a single framework, allowing all operators to contribute on equal footing in a fully model-independent way and enabling consistent connections to generic UV models. The resulting likelihood depends smoothly on the Wilson coefficients and, where applicable, includes Wilson-coefficient–dependent theoretical uncertainties. By formulating the complete likelihood as a differentiable function of the Wilson coefficients, we overcome the practical limitations of computationally expensive sampling and profiling methods and enable the use of modern gradient-based inference techniques in high-dimensional EFT parameter spaces.

More explicitly, the EFT likelihood is constructed as a function of the Wilson coefficients $\vec C(\mu)$ defined at a given renormalization scale $\mu$. For each observable $O_i$, RG evolution to its characteristic scale $\mu_i$ and, where required, matching between EFTs are encoded in evolution matrices $U(\mu_i,\mu)$, yielding the Wilson coefficients relevant for the observable predictions,
\begin{equation}
    \vec C(\mu_i) = U(\mu_i,\mu)\, \vec C(\mu)\,.
\end{equation}
Theoretical predictions for the set of observables $O_i$ are then given by observable expressions $E_i$, which are smooth functions of a set of polynomials $P_k(\vec C(\mu_i))$ in the Wilson coefficients at the corresponding scales,
\begin{equation}
O^{\mathrm{th}}_i(\vec C(\mu_i)) = E_i\!\bigl(P_k(\vec C(\mu_i))\bigr)\,.
\end{equation}
Experimental information is incorporated through likelihood functions for the measured observables, yielding a global likelihood
\begin{equation}
\mathcal{L}(\vec C) = \prod_i \mathcal{L}_i\!\left(O_i^{\mathrm{th}}(\vec C) \,\middle|\, O_i^{\mathrm{exp}}\right),
\end{equation}
where the individual likelihoods $\mathcal{L}_i$ may be correlated and, where applicable, include Wilson-coefficient–dependent theoretical uncertainties and correlations. By construction, $\mathcal{L}(\vec C)$ is a smooth, differentiable function of the Wilson coefficients across the full EFT parameter space.

In practice, our construction relies on a modular representation of the different components of the EFT likelihood. Theory predictions and their dependence on Wilson coefficients are encoded using the recently introduced Polynomial Observable Prediction exchange format (\texttt{POPxf})~\cite{Brivio:2025mww}, which provides a compact and efficient representation of EFT observables and is used here for the first time in a phenomenological application. RG evolution is implemented in terms of evolution matrices connecting Wilson coefficients at different scales. We make the corresponding RG infrastructure and differentiable likelihood framework publicly available through two new packages developed in the context of the present work: \texttt{rgevolve}~\cite{rgevolve}, which provides the evolution matrices, and \texttt{jelli}~\cite{jelli} (JAX-based EFT Likelihoods), which implements the full differentiable likelihood framework and is designed for reuse in a wide range of EFT phenomenology applications. Further technical details and documentation are provided in the \SupplementalMaterial.

\section{Applications of Differentiable EFT Likelihoods}\label{sec:examples}

Differentiability of the EFT likelihood enables inference methods that are difficult or impractical with conventional implementations. First, analytic gradients enable efficient gradient-based optimization, which we use to determine likelihood modes and to perform profiling in frequentist analyses. Second, the analytic Hessian provides direct access to local curvature information, allowing controlled Gaussian approximations and serving as a powerful preconditioner for high-dimensional sampling. Third, gradient information is essential for Hamiltonian Monte Carlo (HMC)~\cite{Duane:1987de}, enabling scalable Bayesian posterior sampling in high-dimensional parameter spaces. In the following, we illustrate these capabilities in progressively more demanding scenarios.

\subsection{Six-dimensional \boldmath{$b \to s\ell\ell$} likelihood}

To demonstrate the practical impact of differentiability in a controlled setting, we first consider a six-dimensional WET likelihood at the renormalization scale $\mu=4.8\,\text{GeV}$, constructed from the $b \to s\ell\ell$ dataset as implemented in Ref.~\cite{Greljo:2022jac}, consisting of 181 observables.
We vary the real parts of the WET Wilson coefficients
\begin{equation}
\label{eq:WET6D}
\{C_{9}^{bs\mu\mu},\, C_{10}^{bs\mu\mu},\, C_{9}^{\prime\,bs\mu\mu},\, C_{10}^{\prime\,bs\mu\mu},\, C_{9}^{bsee},\, C_{10}^{bsee}\}\,,
\end{equation}
as defined in~\cite{Greljo:2022jac}, and construct the full likelihood as described in the previous section.

Using analytic gradients, we determine the likelihood mode with the limited-memory Broyden--Fletcher--Goldfarb--Shanno (L-BFGS) optimization algorithm~\cite{Broyden:1970hba,Fletcher:1970kfz,Goldfarb:1970zio,Shanno:1970qll,Liu:1989esw}. The analytic Hessian at the mode provides the local curvature and enables a controlled Gaussian approximation of the likelihood. In parallel, we perform HMC sampling using the No-U-Turn Sampler (NUTS)~\cite{JMLR:v15:hoffman14a}.
Because the likelihood is differentiable, gradients and Hessians are evaluated exactly and efficiently.

\begin{figure}[t]
        \includegraphics[scale=0.8]{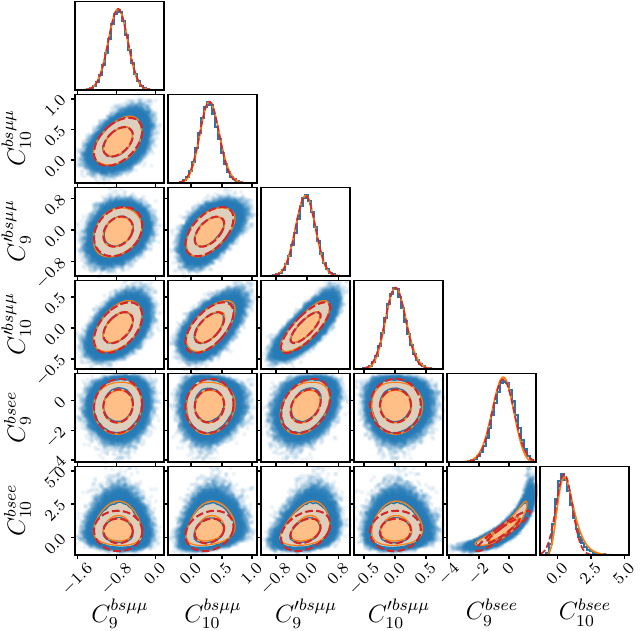}
    \caption{One- and two-dimensional constraints on the WET Wilson coefficients of Eq.~\eqref{eq:WET6D}. The diagonal panels display the 1D marginal posteriors (blue), and Hessian-approximated (red) and profile (orange) likelihoods. The off-diagonal panels show the HMC samples together with the 68\% and 95\% credible regions (blue), the corresponding Gaussian contours from the Hessian at the mode (red), and the 68\% and 95\% confidence regions from the profile likelihoods (orange).}
    \label{fig:bsll}
\end{figure}

Figure~\ref{fig:bsll} compares one- and two-dimensional constraints obtained from (i)~profiling with L-BFGS, (ii)~marginalized HMC samples, and (iii)~the Gaussian approximation derived from the Hessian at the mode. We observe very good agreement for the well-constrained muonic coefficients. Deviations from Gaussianity appear mainly in the electron coefficients, which are less tightly constrained, while the agreement between profiling and marginalization remains excellent. This example illustrates that differentiability allows one to obtain consistent frequentist and Bayesian constraints, as well as controlled Gaussian approximations, within a single computationally efficient framework.

\subsection{A 374-parameter SMEFT fit to Drell--Yan data}\label{sec:DY-only}

We next turn to a high-dimensional SMEFT example based on the neutral- and charged-current $pp\to\ell \ell$ and $pp\to\ell\nu$ Drell--Yan (DY) dataset implemented in Refs.~\cite{Greljo:2022jac,Greljo:2023bab}, with $\ell=e,\mu$, corresponding to 691 observables in total (counting each bin in the binned distributions as an individual observable).

In the SMEFT predictions, we consider all quark flavors present in the parton distribution functions (PDFs) of the initial-state protons.
The absence of the top quark in the PDFs implies unconstrained flat directions if all flavor indices are varied independently. The flavor indices of right-handed quarks can be defined unambiguously in the mass basis, and we exclude Wilson coefficients involving right-handed top quarks as they are unconstrained in our fit. In contrast, no unique mass basis exists for left-handed quark doublets due to CKM misalignment. Excluding specific flavor indices would therefore introduce spurious basis and model dependence.
To obtain basis- and model-independent constraints, we extend the DY likelihood by including four $b\to s\bar\nu\nu$ observables in the fit.
Through $SU(2)_L$ invariance of the SMEFT likelihood, these additional observables remove the remaining flat directions associated with left-handed $\bar t t \bar\ell\ell$ operators.

We construct the SMEFT likelihood at the renormalization scale $\mu=1\,\text{TeV}$ in the Warsaw basis~\cite{Grzadkowski:2010es} and include RG evolution and matching to the WET to connect to the $b\to s\bar\nu\nu$ predictions evaluated at $\mu_{bs\nu\nu}=4.8\,\text{GeV}$.
We simultaneously vary the real and imaginary parts of all dimension-six SMEFT two-quark-two-lepton Wilson coefficients that contribute to the DY and $b\to s\bar\nu\nu$ observables, corresponding to 374 real parameters in total. To our knowledge, this constitutes the highest-dimensional SMEFT fit performed to date.
We use NUTS to sample the 374-dimensional posterior, and we obtain additional one-dimensional profile likelihoods using L-BFGS.

Sampling efficiency depends strongly on the geometry and parameterization of the likelihood.
To enable efficient sampling in this high-dimensional space, we construct a preliminary
linear coordinate transformation based on the exact analytic Hessian $H$ at the Standard Model point.
We perform its eigendecomposition
\begin{equation}
H = Q \Lambda Q^T\,,
\end{equation}
where $Q$ is orthogonal and $\Lambda$ is diagonal.
An orthogonal rotation with $Q^T$ removes local correlations, while a subsequent rescaling
with $|\Lambda|^{1/2}$ normalizes the characteristic scales along each eigendirection.
Defining the transformation matrix
\begin{equation}
K_{\rm HN} = |\Lambda|^{1/2} Q^T\,,
\end{equation}
we introduce Hessian-normalized (HN) coordinates
\begin{equation}
\vec z = K_{\rm HN}\, \vec C\,,
\end{equation}
in which the Hessian at the Standard Model point becomes diagonal with eigenvalues $\pm 1$.
In these HN coordinates, NUTS sampling of the EFT likelihood becomes stable and efficient even in hundreds of dimensions.

While HN coordinates remove local correlations at the Standard Model point, sampling efficiency can be further improved in whitened coordinates, in which the sample covariance matrix is approximately the identity matrix.
We perform a pilot NUTS run in HN coordinates and compute
the sample covariance matrix $\Sigma$. From its eigendecomposition
\begin{equation}
\Sigma = U D U^T\,,
\end{equation}
we construct a whitening transformation
\begin{equation}
K_{\rm W} = D^{-1/2} U^T\,,
\end{equation}
and define whitened coordinates
\begin{equation}
\vec y = K_{\rm W}\, \vec z\,,
\end{equation}
in which the transformed $\Sigma$ is, by construction, the identity matrix, i.e.\ $K_{\rm W}\Sigma K_{\rm W}^T = I$.
In our independent production run in these coordinates, sampling is particularly efficient, and the resulting sample covariance is close to the identity matrix up to finite-sample fluctuations.
The samples in whitened coordinates are finally transformed back into the Warsaw-basis Wilson coefficient coordinates~$\vec C$.

\begin{figure}[t]
        \includegraphics[scale=0.86]{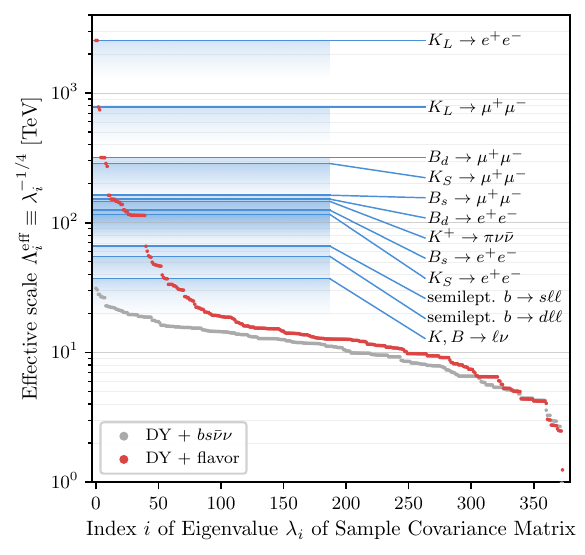}
    \caption{Spectrum of sample-covariance eigenvalues for the DY+$bs\bar\nu\nu$ setup (gray) and for the fit including the full set of low-energy flavor observables (red), expressed in terms of effective scales $\Lambda_i^{\rm eff}\equiv \lambda_i^{-1/4}$, where $\lambda_i$ denotes the corresponding covariance eigenvalue. The horizontal reference lines indicate the highest effective scales at which individual observables or classes of observables provide the dominant impact.}
    \label{fig:spectrum}
\end{figure}

The resulting posterior exhibits strongly non-Gaussian features. Interference terms between Standard Model and new-physics amplitudes linear in certain Wilson coefficients can be compensated by quadratic contributions from many orthogonal directions, generating extended hyperellipsoidal structures in parameter space. In high dimensions, the large volume of these structures shifts the typical posterior region away from the likelihood mode, leading to substantial differences between profiled and marginalized constraints. A detailed analytic explanation of this geometry is provided in the \SupplementalMaterial. All one-dimensional profiled and marginalized bounds in both the up- and down-aligned flavor bases are available as dataset~\cite{dataset} and contained in the \SupplementalMaterial.
A compact summary of the global constraint pattern is provided by the spectrum of $\vec C$-coordinate sample-covariance eigenvalues $\lambda_i$, shown in gray in Fig.~\ref{fig:spectrum}, where we express the constraint on each eigendirection in terms of an effective scale $\Lambda_i^{\rm eff}\equiv \lambda_i^{-1/4}$.
A substantial fraction of eigendirections are already constrained at effective scales above $10\,\text{TeV}$, while the low-scale tail identifies the combinations of Wilson coefficients that remain comparatively weakly bounded.

Importantly, due to the combination of DY and $b\to s\bar\nu\nu$ data, no flavor alignment assumptions have to be imposed a priori, and the full flavor structure can be retained in the posterior samples.
These samples can be transformed into different bases via invertible linear transformations of the likelihood parameters, reflecting the basis independence of the posterior distribution.
While marginalized one-dimensional bounds are inherently basis dependent, they can be obtained in \emph{any} basis from samples of the basis-independent high-dimensional posterior.

\subsection{Multi-scale combination with flavor observables}

Finally, we extend the above DY SMEFT setup by incorporating all low-energy flavor observables that have been used in the likelihood constructions of Refs.~\cite{Greljo:2022jac,Greljo:2023bab} and are sensitive to two-quark-two-lepton operators.
This combination of DY and low-energy data, based on SMEFT and WET predictions evaluated at various scales, fully exploits the multi-scale structure of the likelihood. The total number of included observables is 954.

Compared to the DY+$bs\bar\nu\nu$ setup, the inclusion of the additional flavor observables substantially modifies the likelihood geometry. The relatively homogeneous hyperellipsoidal structures characteristic of the DY likelihood are distorted by the interplay of numerous low-energy constraints, leading to strongly anisotropic curvature and lifting degeneracies along several flavor directions. While some parameter combinations become constrained by orders of magnitude more tightly, others remain comparatively weakly bounded, resulting in a more intricate and less symmetric likelihood geometry. Sampling therefore becomes more challenging and computationally more demanding, but remains feasible in the full 374-dimensional parameter space using the Hessian-based preconditioning strategy described above.

The restructuring of the likelihood geometry is summarized in Fig.~\ref{fig:spectrum}, which shows the spectrum of sample-covariance eigenvalues of the DY+flavor fit in red. Compared to the DY+$bs\bar\nu\nu$ fit, the inclusion of the full set of flavor observables shifts a large fraction of eigenvalues to substantially higher effective scales, in many cases by more than an order of magnitude, while leaving a tail of more weakly constrained directions. The horizontal reference lines indicate the highest effective scales at which individual observables or classes of observables provide the dominant impact, showing how different sectors of the flavor likelihood shape the global constraint pattern. Complete fit results, including all one-dimensional profiled and marginalized bounds in both the up- and down-aligned flavor bases, are available as dataset~\cite{dataset} and contained in the \SupplementalMaterial.

\section{Outlook and Implications}\label{sec:outlook}

In this work, we have introduced differentiable, multi-scale EFT likelihoods and shown that they enable efficient optimization, analytic gradients and local curvature information, and scalable HMC sampling and gradient-based profiling in parameter spaces ranging from a six-dimensional WET example to 374-parameter SMEFT fits. The examples presented here demonstrate that differentiable EFT likelihoods make scalable, basis-independent, and fully multi-scale SMEFT analyses feasible in practice.

Our examples are primarily intended to showcase the power of the approach and can be extended in several directions. On the flavor side, additional observables such as $D$-meson decays~\cite{Fuentes-Martin:2020lea} and charged-lepton-flavour-violating processes can provide complementary constraints on the same class of SMEFT Wilson coefficients. On the collider side, extending the Drell--Yan implementation to include $pp\to\tau\tau$ and $pp\to\tau\nu$ channels~\cite{Faroughy:2016osc, Allwicher:2022mcg, Allwicher:2022gkm} within the same \texttt{POPxf} framework would enlarge the set of simultaneously constrained SMEFT two-quark-two-lepton coefficients. More generally, the examples presented here should not be viewed as providing the strongest possible bounds, but as proof-of-principle demonstrations of a framework that can systematically incorporate additional observables.

Going beyond the two-quark--two-lepton operators probed by Drell--Yan, the inclusion of electroweak precision tests, Higgs observables, dijet processes, and other sectors of observables would allow the construction of genuinely global differentiable SMEFT likelihoods, extending the program initiated in Ref.~\cite{Aebischer:2018iyb}. Likewise, combining with other publicly available likelihood implementations~\cite{Ellis:2020unq, Giani:2023gfq, vanDyk:2021sup, DeBlas:2019ehy, Allwicher:2022mcg}
would further enlarge the scope of the framework, although most existing implementations rely on flavor assumptions that would need to be reconsidered in a model- and basis-independent setup.

While we have focused here on global parameter inference in the SMEFT and the WET, we do not view inference directly in the Wilson-coefficient space as the final goal of the method. Constraints on EFT parameters are most useful as an intermediate step in the interpretation of possible deviations from the Standard Model. The main power of the present framework is that it can be extended straightforwardly to UV models matched to the SMEFT, without imposing \emph{a priori} assumptions at the likelihood level. Combining differentiable matching expressions with differentiable EFT likelihoods would allow the construction of fully differentiable likelihoods for new-physics models. This is considerably more powerful than inference directly in the SMEFT, since UV model parameter spaces are typically much smaller than the corresponding EFT parameter spaces and can therefore be explored much more completely. More fundamentally, the SMEFT is only a low-energy description and must ultimately be UV completed. Only an interpretation in terms of specific UV models can identify the underlying new degrees of freedom and thereby provide concrete targets for future searches for new particles. Such an extension becomes feasible in combination with modern automatic matching tools~\cite{Carmona:2021xtq,Fuentes-Martin:2022jrf}. Fully differentiable model likelihoods based on the present framework and the \texttt{Matchete} matching tool~\cite{Fuentes-Martin:2022jrf} will be investigated in future work.
Moreover, the present setup can be extended naturally to next-to-leading order precision: recent results for higher-order renormalization group evolution in the WET~\cite{Aebischer:2025hsx,Naterop:2025cwg} and in the SMEFT~\cite{Born:2026xkr,Banik:2025wpi}, combined with one-loop matching~\cite{Dekens:2019ept} and NLO predictions for observables, open the way to global NLO analyses of UV models within a fully differentiable framework.

We make the methods developed here publicly available through the accompanying \texttt{jelli}~\cite{jelli} and \texttt{rgevolve}~\cite{rgevolve} packages, facilitating their application and further development in future studies. Their applicability is not restricted to EFT analyses; an immediate further application is the inference of hadronic form-factor parameters and models of non-factorizable hadronic effects in flavor physics~\cite{Altmannshofer:2026cwk}.

More broadly, differentiable likelihoods naturally open the way to information-geometric analyses. In particular, the Fisher information metric can be computed efficiently at arbitrary parameter points, allowing EFT likelihoods to be studied in the context of sloppy models~\cite{Quinn:2021uyo}. For UV models matched to EFTs, this also suggests new information-geometric approaches to questions of fine-tuning and naturalness~\cite{Halverson:2026cxe}.

Differentiable EFT likelihoods therefore do more than improve the efficiency of existing analyses. They define a new framework for precision phenomenology in which global inference, multi-scale consistency, basis independence, and the connection to UV physics can be treated within a single differentiable structure.
We expect this framework to play a central role in the next generation of global analyses of precision data and, more broadly, in the phenomenology of physics beyond the Standard Model.

\section*{Acknowledgments}
Part of this work was carried out at the Mainz Institute for Theoretical Physics (MITP) during the workshop “SMEFT-Tools 2025”, and at the CERN Theoretical Physics Department. We gratefully acknowledge their support and hospitality.
A.S.~acknowledges the financial support from the Slovenian Research and Innovation Agency (Grant No.~J1-50219 and research core funding No.~P1-0035).

\bibliography{references}
\bibliographystyle{JHEP}

\clearpage
\onecolumngrid
\section*{Supplemental Material}
\setcounter{secnumdepth}{3}
\setcounter{section}{0}
\setcounter{subsection}{0}
\setcounter{subsubsection}{0}
\setcounter{equation}{0}
\setcounter{figure}{0}
\setcounter{table}{0}

\renewcommand{\thesection}{S\arabic{section}}
\renewcommand{\thesubsection}{S\arabic{section}.\arabic{subsection}}
\renewcommand{\thesubsubsection}{S\arabic{section}.\arabic{subsection}.\arabic{subsubsection}}

\renewcommand{\theequation}{S\arabic{equation}}
\renewcommand{\thefigure}{S\arabic{figure}}
\renewcommand{\thetable}{S\arabic{table}}

\makeatletter
\renewcommand{\p@section}{}
\renewcommand{\p@subsection}{}
\renewcommand{\p@subsubsection}{}
\makeatother

\begin{center}
\textbf{Table of contents}
\end{center}
\supplementaltableofcontents

\section{Accompanying public tools}
\addsuppsection{Accompanying public tools}

As part of this work, we have developed two new computational tools that we make publicly available: \texttt{jelli}: JAX-based EFT Likelihoods, and \texttt{rgevolve}: Renormalization Group Evolution Matrices for the SMEFT and the WET. We discuss the details of their implementation in the following sections.

\subsection{\texttt{jelli}: JAX-based EFT Likelihoods}
\addsuppsubsection{\texttt{jelli}: JAX-based EFT Likelihoods}

\texttt{jelli}~\cite{jelli} is a Python package for constructing and evaluating likelihood functions in the Effective Field Theory~(EFT) framework, with support for EFTs such as the Standard Model Effective Field Theory (SMEFT) and the Weak Effective Theory (WET). It is designed to be flexible and modular, allowing users to incorporate arbitrary observable predictions provided in the POPxf data format~\cite{Brivio:2025mww} and to accommodate a wide variety of experimental likelihood assumptions. Through its interface with \texttt{rgevolve}~\cite{rgevolve}, discussed in Sec.~\ref{sup:rgevolve}, the framework enables efficient multi-scale analyses via fast renormalization group evolution using the evolution matrix formalism. Built on top of \texttt{JAX}~\cite{jax2018github}, \texttt{jelli} benefits from high-performance numerical computing, automatic differentiation, and Just-In-Time~(JIT) compilation, making the likelihood functions fully differentiable and well suited for gradient-based optimization, sampling, and Hessian calculations. This design enables full likelihood evaluations on the order of milliseconds on a standard laptop and naturally scales to highly parallelized execution on accelerators such as GPUs. The source code is publicly available at \url{https://github.com/jelli-pheno}, while the documentation can be found at \url{https://jelli-pheno.github.io/}.

To infer constraints on the parameter space of an EFT using a set of observables $\vec{O}$, several key ingredients are required. First, one needs theoretical predictions $\vec{O}_\mathrm{th}(\vec{C}, \vec{\theta})$ together with their associated uncertainties, where $\vec{C}$ denotes the Wilson coefficients (the parameters of interest) and $\vec{\theta}$ represents theory nuisance parameters. Second, the framework must account for the renormalization group evolution and matching between different EFTs in order to consistently relate physics across energy scales. Finally, experimental information enters through measurements that define simplified experimental likelihoods, which depend on the theoretical predictions $\vec O_{\mathrm{th}}$, i.e. $\mathcal{L}_{\mathrm{exp}}(\vec O_{\mathrm{th}}) = \mathcal{L}_{\mathrm{exp}}(\vec O_{\mathrm{th}} \mid \vec O_{\mathrm{exp}})$, with experimental nuisance parameters already profiled or marginalized over. Together, these components allow the construction of a global likelihood function that can in general be written as
\begin{equation}
    \mathcal L (\vec C, \vec\theta)= \prod_i \mathcal{L}_\mathrm{exp}^i\!\left(\vec O_{\mathrm{th}} (\vec C, \vec \theta)\mid \vec O_{\mathrm{exp}} \right)\times \mathcal L_\theta(\vec \theta) \,,
\end{equation}
where $i$ labels individual measurements that can constrain arbitrarily many observables, and $\mathcal{L}_\theta(\vec \theta)$ is a distribution constraining the theory nuisance parameters. In a global setting, dealing with a large number of nuisance parameters either by means of marginalization or profiling proves to be a computationally expensive task. In this work, we follow the procedure presented in Refs.~\cite{Aebischer:2018iyb, Stangl:2020lbh} to obtain an approximate nuisance-free likelihood. Namely, we split the observables into two categories:
\begin{enumerate}
    \item Observables with negligible theory uncertainties compared to the experimental ones.
    \item Observables where the theory and experimental uncertainties are both approximated as Gaussian.
\end{enumerate}
The nuisance-free likelihood then factorizes according to the above categories as
\begin{equation}
\label{eq:LikelihoodFactorized}
    \mathcal L (\vec C) = \prod_{i\in 1} \mathcal L_\mathrm{exp}^i\!\left(\vec{O}_\mathrm{th} (\vec C, \vec\theta_0) \mid \vec{O}_\mathrm{exp}\right) \prod_{i \in 2} \tilde{\mathcal L}_\mathrm{exp}^i\!\left(\vec{O}_\mathrm{th}(\vec C, \vec\theta_0)\mid\vec{O}_\mathrm{exp}\right) \,,
\end{equation}
where $\vec\theta_0$ are the central values of the theory nuisance parameters. The first category allows for (in general) non-Gaussian experimental likelihoods, while the second part has the form
\begin{equation}
\label{eq:LikelihoodWithTheory}
-2\ln \tilde{\mathcal{L}}_\mathrm{exp} = \vec{D}^T \left(\Sigma_\mathrm{exp} + \Sigma_\mathrm{th}(\vec{C})\right)^{-1} \vec{D}\,,
\end{equation}
where $\vec{D} = \vec{O}_\mathrm{th} - \vec{O}_\mathrm{exp}$ and $\Sigma_\mathrm{exp}$ and $\Sigma_\mathrm{th}$ are experimental and theoretical covariance matrices.
We extend the setup of Ref.~\cite{Aebischer:2018iyb} by considering the dependence of the theory covariance matrix $\Sigma_\mathrm{th}$ on the Wilson coefficients $\vec{C}$ following the formalism of~\cite{Altmannshofer:2021qrr}.
Below, we discuss each building block of the global likelihood in more detail.

\paragraph{Theory predictions and uncertainties.}
We use the Polynomial Observable Prediction exchange format (\texttt{POPxf}) as the basis for the implementation of theory predictions and the treatment of theory uncertainties and correlations, see Ref.~\cite{Brivio:2025mww} for details. The theory predictions for all observables are expressed as functions of polynomials in the Wilson coefficients, which can be efficiently evaluated in \texttt{JAX}. This allows us to also compute the theoretical covariance matrix $\Sigma_{th}(\vec C)$ as a function of the parameters $\vec C$ from the the covariance matrix of the polynomial coefficients, as described in detail in Ref.~\cite{Brivio:2025mww}, see also~\cite{Altmannshofer:2021qrr}. Having obtained $\Sigma_\mathrm{th} (\vec{C})$, the combined covariance matrix $\Sigma(\vec{C}) \equiv \Sigma_\mathrm{exp} + \Sigma_\mathrm{th} (\vec{C})$ can be computed, the inverse of which enters Eq.~\eqref{eq:LikelihoodWithTheory}.
It is often a good approximation to neglect the dependence of $\Sigma$ on the Wilson coefficients, i.e.~using the constant $\Sigma(\vec 0)$. In this case, the inverse of $\Sigma(\vec{0})$ can be precomputed, and the evaluation of Eq.~\eqref{eq:LikelihoodWithTheory} for different values of $\vec C$ is numerically very efficient. In the general case, however, the inverse of $\Sigma$ depends on $\vec C$ and cannot be precomputed.
To avoid the computationally expensive inversion of $\Sigma$ at each parameter point at which we wish to evaluate the likelihood,
we note that Eq.~\eqref{eq:LikelihoodWithTheory} can be written as
\begin{equation}
    -2\ln \tilde{\mathcal{L}}_\mathrm{exp} = \vec{D}^T\, \vec{x}\,,
\end{equation}
where the vector $\vec{x}$ is defined as $\vec{x}\equiv \Sigma^{-1} \vec{D}$.
This means $\vec x$ is the solution of the system of linear equations
\begin{equation}
\label{eq:ChoSolve}
\Sigma\, \vec{x} = \vec{D} \,,
\end{equation}
which can be solved numerically in a particularly efficient way without having to invert $\Sigma$.
To this end, we use the fact that $\Sigma$ is a symmetric, positive definite matrix and thus admits the Cholesky decomposition~\cite{Cholesky}
\begin{equation}
\Sigma = L L^T\,,
\end{equation}
where $L$ is a real lower triangular matrix with positive diagonal entries.
Consequently, Eq.~\eqref{eq:ChoSolve} can be split into two equations involving only triangular matrices,
\begin{equation}
L\, \vec y = \vec{D}\,, \qquad L^T\, \vec{x} =\vec y \,,
\end{equation}
which are solved by simple forward and back substitution.
Computing the Cholesky decomposition of $\Sigma$ and solving by forward and back substitution is numerically much more efficient than inverting $\Sigma$.
Consequently, the likelihood in Eq.~\eqref{eq:LikelihoodWithTheory} can be evaluated efficiently even when accounting for the dependence of $\Sigma$ on the Wilson coefficients, without requiring a numerically expensive inversion of $\Sigma$ at each parameter point.

\paragraph{RG evolution and matching.}
\texttt{jelli} is interfaced with \texttt{rgevolve} for fast renormalization group evolution using the evolution matrix formalism. See Sec.~\ref{sup:rgevolve} for details.

\paragraph{Experimental likelihoods.}
The final ingredient for defining the likelihood in Eq.~\eqref{eq:LikelihoodFactorized} is specifying the simplified experimental likelihood functions $\mathcal L_\mathrm{exp}$. The functional form
depends on the measurement of a given observable or set of observables
and can be one- or multi-dimensional, allowing for experimental correlations between different observables. \texttt{jelli} contains an efficient \texttt{JAX}-friendly statistics module that supports a broad range of
likelihood distributions, including uni- and multivariate Gaussian likelihoods, distributions for upper limits, likelihoods for Poisson-distributed data, and
distributions given in a purely numerical form.

\subsection{\texttt{rgevolve}: Renormalization Group Evolution Matrices for the SMEFT and the WET}
\addsuppsubsection{\texttt{rgevolve}: Renormalization Group Evolution Matrices for the SMEFT and the WET}
\label{sup:rgevolve}

\texttt{rgevolve}~\cite{rgevolve} is a Python package for fast
renormalization group evolution of Wilson coefficients in the SMEFT and the WET using the evolution matrix formalism. The package provides precomputed evolution, translation, and matching matrices, obtained using \texttt{wilson}~\cite{Aebischer:2018bkb} and \texttt{WCxf}~\cite{Aebischer:2017ugx}. In particular, currently one loop RG evolution in the SMEFT~\cite{Jenkins:2013zja, Jenkins:2013wua, Alonso:2013hga}, the tree-level matching of SMEFT to WET~\cite{Jenkins:2017jig}, and one loop RG evolution in WET~\cite{Aebischer:2017gaw, Jenkins:2017dyc} are implemented, allowing it to be used in both the \texttt{Warsaw} and \texttt{Warsaw up} bases in the SMEFT and \texttt{flavio} and \texttt{JMS} bases in the WET as defined by WCxf~\cite{Aebischer:2017ugx}, with plans to implement the known results at higher loop orders~\cite{Dekens:2019ept,Aebischer:2025hsx,Naterop:2025cwg,Born:2026xkr,Banik:2025wpi} in future releases. The source code of the package can be found at \url{https://github.com/rgevolve}, including an up-to-date list of the available EFTs and bases.

The RG evolution matrix $R(\mu, \mu_0)$ allows expressing the solution of a linear renormalization group equation of the Wilson coefficients $\vec C(\mu)$ with anomalous dimension matrix $\gamma(\mu)$,
\begin{equation}
\frac{d}{d \ln \mu} \vec{C} (\mu) = \gamma^T\!(\mu)\, \vec{C}(\mu)
\end{equation}
with boundary condition $\vec C(\mu_0) = \vec C_0$, as
\begin{equation}
    \vec{C}(\mu) = R(\mu, \mu_0) \vec{C}(\mu_0)\,.
\end{equation}
This matrix satisfies the analogous differential equation
\begin{equation}\label{eq:RGE_R}
\frac{d}{d \ln \mu} R(\mu, \mu_0) = \gamma^T\!(\mu)\,  R(\mu, \mu_0)
\end{equation}
and the boundary condition $R(\mu_0, \mu_0)= I$, with $I$ the identity matrix.

Matching and translation matrices $M$ and $T$ that relate Wilson coefficients $\vec{C}$ in one EFT or basis to Wilson coefficients $\vec C'$ in another EFT or basis, can be computed at a fixed renormalization scale, so that $\vec{C}^\prime (\mu) = M(\mu) \vec{C} (\mu)$, and $\vec{C}^\prime (\mu) = T(\mu) \vec{C} (\mu)$.

A combined evolution of Wilson coefficients, involving matching and/or translation, can then be written as a series of matrix products, combining matrices $R$, $M$, and $T$. Denoting this combination as $U$, we can write the complete map from Wilson coefficients $\vec C$ at scale $\mu$ in a given EFT to the Wilson coefficients $\vec C'$ entering an observable prediction in a potentially different EFT or basis at scale $\mu_i$ as
\begin{equation}
    \vec{C}^\prime (\mu_i) = U(\mu_i, \mu) \vec{C}(\mu) \,.
\end{equation}
By precomputing these matrices for various scales $\mu$ and $\mu_i$, RG evolution, matching, and basis translation can be implemented in a numerically efficient way.

The smooth dependence of the RG evolution on the scale can be exploited to obtain approximate results for scales at which the evolution matrices have not been explicitly precomputed.
To be specific,
given two precomputed matrices $R(\mu,\Lambda_1)$ and $R(\mu,\Lambda_2)$,
we use the fact that the solution of Eq.~\eqref{eq:RGE_R} leads to an approximately logarithmic scale dependence for small scale separation and
approximate the evolution matrix at an intermediate scale $\Lambda$ with $\Lambda_1 < \Lambda < \Lambda_2$ by interpolating in the logarithm of the scale,
\begin{equation}
R(\mu,\Lambda) \approx
R(\mu,\Lambda_1)
+
\left(
R(\mu,\Lambda_2) - R(\mu,\Lambda_1)
\right)
\frac{\log(\Lambda/\Lambda_1)}{\log(\Lambda_2/\Lambda_1)}\,.
\end{equation}
This interpolation provides an efficient approximation that allows evolving Wilson coefficients to an arbitrary intermediate scale without requiring
explicitly solving Eq.~\eqref{eq:RGE_R}.
The resulting interpolation error depends on the separation between $\Lambda_1$ and $\Lambda_2$ and can be systematically reduced by precomputing evolution matrices at a denser set of scales.

\section{Detailed results of the fits}
\addsuppsection{Detailed results of the fits}
\subsection{Six-dimensional $b\to s \ell \ell$ fit}
\addsuppsubsection{Six-dimensional $b\to s \ell \ell$ fit}

We use the following definition of the weak effective Hamiltonian
\begin{equation}
\mathcal{H}_\mathrm{eff} = \mathcal{H}_\mathrm{eff}^\mathrm{SM} - \frac{4 G_F}{\sqrt{2}} \frac{e^2}{16\pi^2} V_{tb} V_{ts}^\ast\sum_{\ell=e, \mu} \sum_{i}(C_i^{bs\ell\ell} O_i^{bs\ell\ell} + C_i^{\prime bs\ell\ell} O_i^{\prime bs\ell\ell}) + \mathrm{h.c.}\,,
\label{eq:WET}
\end{equation}
where the Wilson coefficients $C_i$ denote the pure NP contributions and are defined at the renormalization scale $\mu=4.8~\mathrm{GeV}$. The semi-leptonic operators considered in this work are defined as
\begin{align}
O_9^{bs\ell\ell} &=
(\bar{s} \gamma_{\mu} P_{L} b)(\bar{\ell} \gamma^\mu \ell)\,,
&
O_9^{\prime bs\ell\ell} &=
(\bar{s} \gamma_{\mu} P_{R} b)(\bar{\ell} \gamma^\mu \ell)\,,\label{eq:O9}
\\
O_{10}^{bs\ell\ell} &=
(\bar{s} \gamma_{\mu} P_{L} b)( \bar{\ell} \gamma^\mu \gamma_5 \ell)\,,
&
O_{10}^{\prime bs\ell\ell} &=
(\bar{s} \gamma_{\mu} P_{R} b)( \bar{\ell} \gamma^\mu \gamma_5 \ell)\,.\label{eq:O10}
\end{align}

\subsubsection{Dataset}
\addsuppsubsubsection{Dataset}

We leverage the large number of flavor observables implemented in the \texttt{flavio}~\cite{Straub:2018kue} package to perform the $b\to s \ell \ell$ fits discussed in this work. Many decays with this underlying quark transition are included, e.g.~the (differential) branching fractions of $B\to K^{(*)} \mu \mu$, $B_s \to \phi \mu \mu$ and $\Lambda_b \to \Lambda \mu \mu$, $B_s \to \mu \mu$, as well as the angular observables and the LFU ratios $R_{K^{(*)}}$. See Refs.~\cite{Greljo:2022jac, Altmannshofer:2021qrr} for more details. With respect to those works, we have updated the $B\to K$ form factors with the results of Ref.~\cite{Gubernari:2023puw}, and we have used a different CKM input scheme based on the unitarity triangle angle $\beta$ from the latest HFLAV determination~\cite{HeavyFlavorAveragingGroupHFLAV:2024ctg} instead of the exclusive determination of $|V_{ub}|$. Altogether our fits include 181 observables, all of which we have expressed as (functions of) polynomials in the contributing Wilson coefficients to produce \texttt{POPxf} files. The corresponding measurement files have been directly obtained from \texttt{flavio}'s database of measurements.

\subsubsection{Results}
\addsuppsubsubsection{Results}

We have considered three approaches to perform the 6D $b\to s \ell \ell$ fit:
\begin{itemize}
    \item A Bayesian analysis: We have sampled the posterior using the NUTS HMC sampler implemented in \texttt{numpyro}~\cite{phan2019composable, bingham2019pyro}. To this end, we have adopted flat, uninformative priors for all parameters. We have used 100 chains initialized at the SM point with different random seeds, each consisting of 1000 warm-up iterations used to adapt the HMC parameters, followed by 2000 sampling iterations. Altogether we have thus obtained 200k samples of high quality based on standard diagnostics, such as the potential scale reduction statistic $\hat{R}$, the effective sample size $n_{\mathrm{eff}}$, and the absence of divergent transitions~\cite{1992StaSc...7..457G, 10.1214/20-BA1221, geyer2011mcmc}.
    \item A frequentist analysis: We have determined 1D and 2D profile likelihoods using the L-BFGS optimization algorithm implemented in \texttt{scipy}~\cite{2020SciPy-NMeth}.
    \item A Gaussian approximation of the likelihood with the mean given by the maximum likelihood estimate (MLE), and the covariance matrix given by the inverse of the Hessian matrix at the MLE. The MLE has been obtained using L-BFGS and the Hessian matrix is directly available due to the differentiability of the likelihood.
\end{itemize}

The summary of 1D marginalized and profiled constraints is given in Tab.~\ref{tab:bsll_constraints}, while the local Gaussian approximation in terms of MLE, standard deviations, and correlation matrix is given in Tab.~\ref{tab:wet_gaussian_summary}. These data, together with the full samples, are available at~\cite{dataset}. We find excellent agreement between the different methods for all the muonic Wilson coefficients, for which the amount of data available implies a close-to-Gaussian global likelihood. For the electron Wilson coefficients less data is available, and we observe a larger discrepancy between the Gaussian approximation and the two other approaches.

\begin{table}
\caption{Summary of marginalized and profiled constraints for the 6D $ b \to s \ell \ell $ fit.}
\label{tab:bsll_constraints}
\centering
\small
\begin{tabular}{@{}>{\raggedright\arraybackslash}p{0.18\linewidth}@{}>{\raggedright\arraybackslash}p{0.27\linewidth}@{}>{\raggedright\arraybackslash}p{0.27\linewidth}@{}>{\raggedright\arraybackslash}p{0.28\linewidth}@{}}
\toprule
Parameter & Posterior mean $\pm$ std & Posterior median (68\% CI) & Profile (68\% CL) \\
\midrule
$C_{9}^{bs\mu\mu}$ & $-0.77 \pm 0.20$ & $-0.76^{+0.20}_{-0.20}$ & $-0.77^{+0.20}_{-0.20}$ \\
\rowcolor{gray!15} $C_{10}^{bs\mu\mu}$ & $0.29 \pm 0.16$ & $0.29^{+0.16}_{-0.16}$ & $0.30^{+0.16}_{-0.16}$ \\
$C_{9}'^{bs\mu\mu}$ & $-0.03 \pm 0.25$ & $-0.03^{+0.25}_{-0.25}$ & $-0.04^{+0.25}_{-0.25}$ \\
\rowcolor{gray!15} $C_{10}'^{bs\mu\mu}$ & $0.01 \pm 0.17$ & $0.00^{+0.17}_{-0.16}$ & $0.00^{+0.17}_{-0.16}$ \\
$C_{9}^{bsee}$ & $-0.31 \pm 0.73$ & $-0.29^{+0.72}_{-0.75}$ & $-0.36^{+0.70}_{-0.74}$ \\
\rowcolor{gray!15} $C_{10}^{bsee}$ & $0.63 \pm 0.68$ & $0.54^{+0.74}_{-0.55}$ & $0.50^{+0.70}_{-0.53}$ \\
\bottomrule
\end{tabular}
\end{table}


\begin{table}[t]
\caption{Best-fit values and correlation matrix obtained from the local Gaussian approximation at the maximum of the likelihood for the 6D $ b \to s \ell \ell $ fit.}
\label{tab:wet_gaussian_summary}
\centering
\makebox[\textwidth][c]{%
\begin{tabular}{@{}lc@{}}
\toprule
Parameter & Best fit $\pm 1\sigma$ \\
\midrule
$C_{9}^{bs\mu\mu}$ & $-0.77 \pm 0.20$ \\
$C_{10}^{bs\mu\mu}$ & $+0.30 \pm 0.16$ \\
$C_{9}'^{bs\mu\mu}$ & $-0.04 \pm 0.25$ \\
$C_{10}'^{bs\mu\mu}$ & $+0.00 \pm 0.17$ \\
$C_{9}^{bsee}$ & $-0.36 \pm 0.73$ \\
$C_{10}^{bsee}$ & $+0.50 \pm 0.61$ \\
\bottomrule
\end{tabular}
\hspace{0.03\textwidth}
\begin{tabular}{@{}lcccccc@{}}
\toprule
 & $C_{9}^{bs\mu\mu}$ & $C_{10}^{bs\mu\mu}$ & $C_{9}'^{bs\mu\mu}$ & $C_{10}'^{bs\mu\mu}$ & $C_{9}^{bsee}$ & $C_{10}^{bsee}$ \\
\midrule
$C_{9}^{bs\mu\mu}$ & 1 & 0.45 & 0.25 & 0.47 & 0.07 & -0.02 \\
$C_{10}^{bs\mu\mu}$ &  & 1 & 0.53 & 0.59 & 0.06 & 0.23 \\
$C_{9}'^{bs\mu\mu}$ &  &  & 1 & 0.79 & 0.31 & 0.34 \\
$C_{10}'^{bs\mu\mu}$ &  &  &  & 1 & 0.00 & 0.06 \\
$C_{9}^{bsee}$ &  &  &  &  & 1 & 0.94 \\
$C_{10}^{bsee}$ &  &  &  &  &  & 1 \\
\bottomrule
\end{tabular}%
}
\end{table}

\subsection{374-dimensional SMEFT fits}
\addsuppsubsection{374-dimensional SMEFT fits}
\begin{table}[t]
\centering
\caption{The semi-leptonic SMEFT operators considered in this work, where $q$ and $l$ are the quark and lepton $SU(2)_L$ doublets, $u,d,e$ are the right handed $SU(2)_L$ singlets, $p,r,s,t$ are the flavor indices, and $j,k$ are the $SU(2)_L$ indices.}
\label{tab:SMEFToperators}
\begin{tabular}{ll}
\toprule
Operator & Definition \\
\midrule
$\mathcal{O}_{lq}^{(1)}$ &
$(\bar l_p \gamma_\mu l_r)(\bar q_s \gamma^\mu q_t)$ \\

$\mathcal{O}_{lq}^{(3)}$ &
$(\bar l_p \gamma_\mu \sigma^i l_r)(\bar q_s \gamma^\mu \sigma^i q_t)$ \\

\midrule
$\mathcal{O}_{lu}$ &
$(\bar l_p \gamma_\mu l_r)(\bar u_s \gamma^\mu u_t)$ \\

$\mathcal{O}_{ld}$ &
$(\bar l_p \gamma_\mu l_r)(\bar d_s \gamma^\mu d_t)$ \\

$\mathcal{O}_{qe}$ &
$(\bar q_p \gamma_\mu q_r)(\bar e_s \gamma^\mu e_t)$ \\

\midrule
$\mathcal{O}_{eu}$ &
$(\bar e_p \gamma_\mu e_r)(\bar u_s \gamma^\mu u_t)$ \\

$\mathcal{O}_{ed}$ &
$(\bar e_p \gamma_\mu e_r)(\bar d_s \gamma^\mu d_t)$ \\

\midrule
$\mathcal{O}_{ledq}$ &
$(\bar l_p^j e_r)(\bar d_s q_{t j})$ \\

$\mathcal{O}_{lequ}^{(1)}$ &
$(\bar l_p^j e_r)\,\varepsilon_{jk}(\bar q_s^k u_t)$ \\

$\mathcal{O}_{lequ}^{(3)}$ &
$(\bar l_p^j \sigma_{\mu\nu} e_r)\,\varepsilon_{jk}(\bar q_s^k \sigma^{\mu\nu} u_t)$ \\
\bottomrule
\end{tabular}
\end{table}

We use the following definition of the SMEFT Lagrangian
\begin{equation}
\mathcal{L_{\mathrm{SMEFT}}} = \mathcal{L_{\mathrm{SM}}} + \sum_{\mathcal{O}_i=\mathcal{O}_i^\dagger} {C_i} \mathcal{O}_i + \sum_{\mathcal{O}_i\neq \mathcal{O}_i^\dagger} \left( {C_i}\mathcal{O}_i + {C_i^\ast} \mathcal{O}_i^\dagger \right) \,
\label{eq:SMEFT}
\end{equation}
where the sums run over the hermitian and non-hermitian operators and $C_i$ are dimensionful Wilson coefficients. The semi-leptonic dimension-six operators considered in this work are defined in Tab.~\ref{tab:SMEFToperators}.

\subsubsection{Dataset}
\addsuppsubsubsection{Dataset}
\label{sup:DYdataset}

Both of the SMEFT fits considered in this work crucially rely on the neutral and charged current Drell--Yan implementation in \texttt{flavio} presented in Ref.~\cite{Greljo:2022jac}, which consists of 691 bins measured across the electron and muon channels by both CMS~\cite{CMS:2021ctt, CMS:2022yjm} and ATLAS~\cite{ATLAS:2019lsy,ATLAS:2020yat}. We consider all the semileptonic operators defined in Tab.~\ref{tab:SMEFToperators} with the following restrictions on the flavor indices. For operators involving only right-handed leptons, we only consider lepton flavor conserving operators. We do allow for lepton flavor violation in the operators involving left-handed leptons, as the neutrino flavor cannot be resolved in charged-current Drell--Yan. On the quark side, Drell--Yan processes are insensitive to operators involving the top quark, since the top is absent from the parton distribution functions of the initial state protons. As the flavor indices of right-handed quarks can be defined unambiguously in the mass basis, we exclude the operators involving right-handed top quarks. On the other hand, no unique mass basis exists for left-handed quarks in the presence of CKM mixing, and thus we retain operators involving the third-generation quark doublet to not introduce basis-dependence. Their presence gives rise to a direction in our parameter space to which Drell--Yan phenomenology is insensitive. This can be understood from the following $SU(2)$ decomposition written in the up-aligned basis
\begin{equation}
\label{eq:lq1-lq3}
\begin{aligned}
{[O_{\ell q}^{(1)}-O_{\ell q}^{(3)}]}_{pr33}
&= 2(\bar e_p\gamma_\mu P_L e_r)(\bar t\gamma^\mu P_L t)
+2V_{t\beta}^*V_{t\alpha}(\bar{\nu}_p\gamma_\mu P_L\nu_r)(\bar d_\beta\gamma^\mu P_L d_\alpha)\\
&-2V_{t\alpha}(\bar e_p\gamma_\mu P_L\nu_r)(\bar t\gamma^\mu P_L d_\alpha) -2V_{t\beta}^*(\bar{\nu}_p\gamma_\mu P_L e_r)(\bar d_\beta\gamma^\mu P_L t)\,.
\end{aligned}
\end{equation}
We emphasize that the presence of a flat direction is basis-independent, and we only demonstrate it in the up-aligned basis for concreteness. To remove the flat direction related to Eq.~\eqref{eq:lq1-lq3}, we include $b\to s \nu \nu$ data in both of our fits, consisting of four branching fractions of $B \to K^{(*)} \nu\nu$, including the latest Belle-II measurement~\cite{Belle-II:2023esi}.
In total the number of real parameters in this setup is 374.

The second fit adds a rich set of flavor observables on top of the Drell--Yan and $b\to s \nu \nu$ data, as discussed and used in Refs.~\cite{Greljo:2022jac, Greljo:2023bab}. This consists of a wide range of observables based on the underlying $b\to s$, $b\to d$, $b\to u$, $s\to d$, $s\to u$ and $d\to u$ transitions, which represent a subset of the current \texttt{smelli}~\cite{Aebischer:2018iyb} dataset. Altogether the DY+flavor fit consists of 954 observables. We have expressed all of them as (functions of) polynomials in the contributing Wilson coefficients to produce \texttt{POPxf} files. The corresponding measurement files have been directly obtained from \texttt{flavio}'s database of measurements.

\subsubsection{Analytic explanation of volume effects in Drell--Yan}
\addsuppsubsubsection{Analytic explanation of volume effects in Drell--Yan}
\label{sup:volume_effects}

The global SMEFT analyses performed in this work explore high-dimensional parameter spaces in which volume effects make it particularly important to distinguish between constraints derived from marginal posterior distributions and profile likelihood constraints. The marginal posterior for a parameter $C_i$ from $\vec{C}$ is obtained by integrating over the remaining parameters,
\begin{equation}
    p(C_i \mid \vec{O}_\mathrm{exp}) \propto \int
\mathcal{L}(\vec{C})
\,\pi(\vec C)\, dC_{j\neq i},
\end{equation}
where $\pi(\vec C)$ denotes the prior. This integration weights regions of parameter space by both their likelihood and their volume. Consequently, extended regions with somewhat lower likelihood can dominate over narrower regions with higher peak likelihood. Since the volume scales with the power of the dimensionality, volume effects become increasingly important in high-dimensional parameter spaces.
Marginal posterior distributions can become dominated by large-volume regions and show spurious tensions with the SM, even though the maximum-likelihood estimate remains close to the SM. This behavior is a generic feature of high-dimensional SMEFT analyses with many operators entering observables only at quadratic order, with few dominant linear directions.
In contrast, profile likelihood constraints are obtained by maximizing the likelihood with respect to the remaining parameters and are therefore insensitive to such effects.

Importantly, the linear dependence of all kinematic bins of the differential Drell--Yan distributions is dominated by the same combination of Wilson coefficients. As a consequence, combining the bins leads to the geometrical structure and the volume effects described below.
In the following, we define $C_{\mathrm{lin}}$ as the linear combination of Wilson coefficients that interferes with the SM and $C_i^\perp$ as the $N_\perp$ perpendicular directions to $C_{\mathrm{lin}}$. In a given kinematic bin~$k$, the cross-section can then be expressed as
\begin{equation}
\sigma_k =
\sigma_k^{\mathrm{SM}}
+ C_{\mathrm{lin}}\,\alpha_k
+ C_{\mathrm{lin}}^2\,\beta^{(0)}_k
+ \sum_{i=1}^{N_\perp} |C_i^\perp|^2\,\beta^{(i)}_k \,,
\end{equation}
where $\alpha, \beta$ are bin-dependent coefficients arising from the interference and quadratic contributions. Since experimentally $\sigma_k^{(exp)}\approx \sigma_k^{\mathrm{SM}}$, one can define a "radius" parameter $R_k$ such that
\begin{equation}
\label{sup:eq:DYvol}
R_k^2(C_{\mathrm{lin}}) \approx  - \alpha_k\,C_{\mathrm{lin}}
- \beta^{(0)}_k\, C_{\mathrm{lin}}^2
\end{equation}
with
\begin{equation}
R_k^2 = \sum_{i=1}^{N_\perp} |C_i^\perp|^2\,\beta^{(i)}_k\,.
\end{equation}
The latter equation defines a hyperellipsoid in an $N_\perp$-dimensional space.
To simplify the discussion, we consider rescaled perpendicular directions
\begin{equation}
    \tilde C^\perp_i = C^\perp_i |\beta_k^{(i)}|^{-1/2}\,,
\end{equation}
in which the hyperellipsoid becomes a hypersphere of radius $R_k$.
Through Eq.~\eqref{sup:eq:DYvol}, the radius of this hypersphere depends on the value of $C_{\mathrm{lin}}$ and vanishes close to the SM point $C_{\mathrm{lin}}=0$. Crucially, the volume of a thin spherical shell in $N_\perp$ dimensions, with its thickness given by the experimental and theoretical uncertainties, scales as $R_k^{N_\perp-1}$. Near the SM point, the radius scales roughly as $R_k^2\propto C_{\mathrm{lin}}$ and the volume increases as we move along $C_{\mathrm{lin}}$ with approximately a factor of $|C_{\mathrm{lin}}|^{(N_\perp-1)/2}$.

Once we combine information from multiple bins, assuming Gaussian likelihoods for simplicity, the combined likelihood is obtained from the intersections of each bin's hypersphere.
This leads to an exponential suppression of the likelihood as we move away from the SM point along $C_{\mathrm{lin}}$, schematically given by $\exp\!\left(-\frac{1}{2}\frac{C_{\mathrm{lin}}^2}{\sigma^2}\right)$.
However, with the increase in volume, the posterior is
\begin{equation}
p(C_{\mathrm{lin}} \mid \vec{O}_\mathrm{exp}) \propto \exp\!\left(-\frac{1}{2}\frac{C_{\mathrm{lin}}^2}{\sigma^2}\right) \times |C_{\mathrm{lin}}|^{(N_\perp-1)/2}\,,
\end{equation}
which peaks at $|C_{\mathrm{lin}}|= \sqrt{(N_\perp-1)/2}\ \sigma \approx 7 \sigma$ for $N_\perp\approx 100$.
The linear combination $C_{\mathrm{lin}}$ thus shows a spurious tension with the SM in the marginal posterior, even though the profile likelihood remains consistent with the SM.

\begin{figure}[t]
        \includegraphics[scale=0.8]{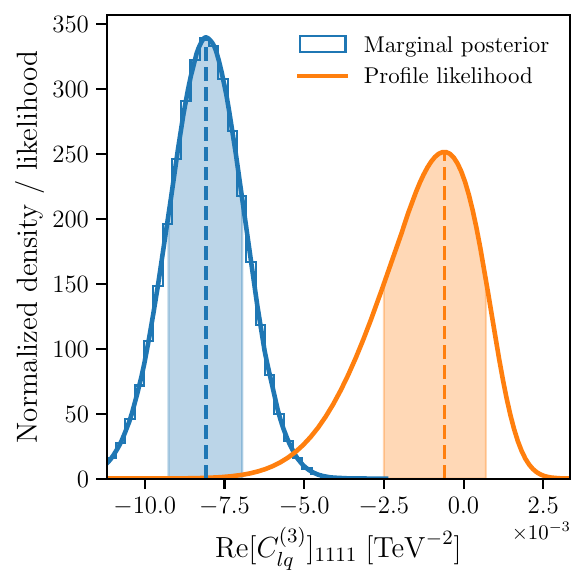}
    \caption{Normalized marginal posterior and profile likelihood for $\mathrm{Re}[C_{lq}^{(3)}]_{1111}$ from the DY+$bs\bar\nu\nu$ analysis. The shaded regions show the 68\% posterior credible interval and the profile likelihood 68\% confidence interval, while the vertical dashed lines indicate the posterior median and the best-fit value, respectively.}
    \label{fig:lq3_1111}
\end{figure}

Several semi-leptonic operators contribute linearly to Drell--Yan production through interference with the SM amplitudes. Among these, the operator $\mathcal{O}_{lq}^{(3)}$ with first-generation quarks generates the dominant interference contribution from the real part of its Wilson coefficient. This is partly due to the enhancement from proton parton luminosities, but also the electroweak structure of the interaction plays an important role: the isospin current appearing in $\mathcal{O}_{lq}^{(3)}$ aligns well with the photon and $Z$ couplings, leading to efficient interference with the SM amplitudes. In contrast, the flavor-singlet structure of $\mathcal{O}_{lq}^{(1)}$ induces interference contributions from up- and down-quark channels with opposite signs, which partially cancel. Operators involving right-handed fermions, such as $\mathcal{O}_{eu}$ or $\mathcal{O}_{lu}$, can also interfere with the SM amplitudes but typically yield smaller effects because the relevant SM neutral-current couplings are weaker.
Consequently, $C_{\mathrm{lin}}$ is dominated by the real part of the Wilson coefficient of $\mathcal{O}_{lq}^{(3)}$ with first-generation quarks, and thus the marginalized constraint on this coefficient is most affected by volume effects.
This is illustrated in Fig.~\ref{fig:lq3_1111}, obtained using the results of the DY+$bs\bar\nu\nu$ fit discussed below, showing a clear difference between the values of $\mathrm{Re}[C_{lq}^{(3)}]_{1111}$ inferred from the marginal posterior and the profile likelihood.

\subsubsection{Results}
\addsuppsubsubsection{Results}

We have conducted two SMEFT analyses in a 374 dimensional parameter space using the datasets discussed in Sec.~\ref{sup:DYdataset}. We have performed HMC sampling of the full posteriors using the NUTS sampler implemented in \texttt{numpyro}~\cite{phan2019composable, bingham2019pyro}, adopting flat, uninformative priors for all parameters. The pilot runs performed in Hessian-normalized coordinates consisted of 100 chains initialized at the SM point using different random seeds, each with 1000 warm-up and 2000 sampling iterations, altogether constituting 200k samples per dataset. From these, we have derived whitened coordinates in which the production sampling was performed. For each dataset, we have run 400 independent chains, each with 1000 warm-up and 5000 sampling iterations, producing 2M samples per dataset. The produced samples are of high quality based on standard diagnostics such as the potential scale reduction statistic $\hat{R}$, the effective sample size $n_{\mathrm{eff}}$, and the absence of divergent transitions~\cite{1992StaSc...7..457G, 10.1214/20-BA1221, geyer2011mcmc}, confirming good mixing and convergence of all chains.
While we have defined the Hessian-normalized and whitened coordinates with respect to the up-aligned Warsaw basis, the resulting samples are basis-independent since the full set of semi-leptonic operators involving left-handed quark doublets has been included. The samples can thus be simply transformed into the down-aligned basis before computing the 1D marginalized constraints.

To obtain 1D profiled constraints, we have used the L-BFGS optimization algorithm implemented in \texttt{scipy}~\cite{2020SciPy-NMeth} to minimize the negative log-likelihood with respect to the remaining 373 directions.
To improve the convergence of the minimization, we have used the HMC samples to derive whitened coordinates in the 373 dimensional space orthogonal to the profiled direction.
The 1D profiled constraints are intrinsically basis-dependent and have to be performed separately for the up-aligned and down-aligned bases.

The full set of one-dimensional profiled and marginalized bounds in both the up- and down-aligned flavor bases are provided in Tab.~\ref{tab:DY_constraints}, for both the DY+$bs\bar\nu\nu$ and DY+flavor datasets. They are also available at~\cite{dataset}, together with the full samples in the up-aligned basis. We observe large volume effects for the Wilson coefficients $\mathrm{Re}[C_{lq}^{(3)}]_{1111}$ and $\mathrm{Re}[C_{lq}^{(3)}]_{2211}$, as expected from the discussion in Sec.~\ref{sup:volume_effects}. This can be seen by comparing the marginalized and profiled constraints: the former show a spurious significant tension with respect to the SM, while the latter are perfectly consistent with the SM.
We observe much milder volume effects also for the Wilson coefficients of the $\mathcal{O}_{eu}$ and $\mathcal{O}_{lu}$ operators involving first-generation up-quarks.


\setlength{\LTleft}{0pt}
\setlength{\LTright}{0pt}
\scriptsize
\renewcommand{\arraystretch}{1.4}
\begin{longtable}{@{}>{\centering\arraybackslash}m{0.150\linewidth}>{\raggedleft\arraybackslash}m{0.110\linewidth}p{0.020\linewidth}>{\raggedleft\arraybackslash}p{0.150\linewidth}>{\raggedleft\arraybackslash}p{0.150\linewidth}p{0.020\linewidth}>{\raggedleft\arraybackslash}p{0.150\linewidth}>{\raggedleft\arraybackslash}p{0.150\linewidth}p{0.020\linewidth}@{}}
\caption{Summary of marginalized and profiled constraints for the 374-dimensional SMEFT fits considered in this work. Each cell shows from top to bottom: the posterior median with the 68\% credible interval and the profile likelihood maximum with the 68\% confidence interval. All Wilson coefficients are given in units of $[\mathrm{TeV}^{-2}]$ and defined at the renormalization scale $\mu=1~\mathrm{TeV}.$}\label{tab:DY_constraints}\\
\toprule
\makecell[c]{Parameter\\ $[\mathrm{TeV}^{-2}]$} &  &  & \multicolumn{2}{c}{up-aligned} &  & \multicolumn{2}{c}{down-aligned} &  \\
 & & & \multicolumn{1}{r}{\makecell{DY + $bs\bar\nu\nu$}} & \multicolumn{1}{r}{\makecell{DY + flavor}} &  & \multicolumn{1}{r}{\makecell{DY + $bs\bar\nu\nu$}} & \multicolumn{1}{r}{\makecell{DY + flavor}} \\
\midrule
\endfirsthead
\toprule
\makecell[c]{Parameter\\ $[\mathrm{TeV}^{-2}]$} &  &  & \multicolumn{2}{c}{up-aligned} &  & \multicolumn{2}{c}{down-aligned} &  \\
 & & & \multicolumn{1}{r}{\makecell{DY + $bs\bar\nu\nu$}} & \multicolumn{1}{r}{\makecell{DY + flavor}} &  & \multicolumn{1}{r}{\makecell{DY + $bs\bar\nu\nu$}} & \multicolumn{1}{r}{\makecell{DY + flavor}} \\
\midrule
\endhead
\midrule
\multicolumn{8}{r}{\emph{Continued on next page}} \\
\midrule
\endfoot
\bottomrule
\endlastfoot
\multirow{2}{*}{$\mathrm{Re}\!\left[C_{ed}\right]_{1111}$} & \makecell[tr]{marginalized: \\[4pt] profiled:} &  & \makecell[tr]{$-2.9^{+6.7}_{-6.8}\times 10^{-3}$ \\[4pt] $-0.1^{+2.9}_{-2.7}\times 10^{-2}$} & \makecell[tr]{$-2.1 \pm 7.5\times 10^{-3}$ \\[4pt] $0.0 \pm 2.1\times 10^{-2}$} &  & \makecell[tr]{$-2.9^{+6.7}_{-6.8}\times 10^{-3}$ \\[4pt] $-0.1^{+2.9}_{-2.7}\times 10^{-2}$} & \makecell[tr]{$-2.1 \pm 7.5\times 10^{-3}$ \\[4pt] $0.0 \pm 2.1\times 10^{-2}$} & \\
\rowcolor{gray!15} \multirow{2}{*}{$\mathrm{Im}\!\left[C_{ed}\right]_{1112}$} & \makecell[tr]{marginalized: \\[4pt] profiled:} &  & \makecell[tr]{$0.0 \pm 8.4\times 10^{-3}$ \\[4pt] $0.0 \pm 3.5\times 10^{-2}$} & \makecell[tr]{$4.3^{+7.8}_{-7.9}\times 10^{-3}$ \\[4pt] $1.2^{+1.6}_{-2.1}\times 10^{-2}$} &  & \makecell[tr]{$0.0 \pm 8.4\times 10^{-3}$ \\[4pt] $0.0 \pm 3.5\times 10^{-2}$} & \makecell[tr]{$4.3^{+7.8}_{-7.9}\times 10^{-3}$ \\[4pt] $1.2^{+1.6}_{-2.1}\times 10^{-2}$} & \\
\multirow{2}{*}{$\mathrm{Re}\!\left[C_{ed}\right]_{1112}$} & \makecell[tr]{marginalized: \\[4pt] profiled:} &  & \makecell[tr]{$0.0 \pm 8.4\times 10^{-3}$ \\[4pt] $0.0 \pm 3.5\times 10^{-2}$} & \makecell[tr]{$6.9^{+8.8}_{-8.9}\times 10^{-3}$ \\[4pt] $1.9^{+1.6}_{-2.6}\times 10^{-2}$} &  & \makecell[tr]{$0.0 \pm 8.4\times 10^{-3}$ \\[4pt] $0.0 \pm 3.5\times 10^{-2}$} & \makecell[tr]{$6.9^{+8.8}_{-8.9}\times 10^{-3}$ \\[4pt] $1.9^{+1.6}_{-2.6}\times 10^{-2}$} & \\
\rowcolor{gray!15} \multirow{2}{*}{$\mathrm{Im}\!\left[C_{ed}\right]_{1113}$} & \makecell[tr]{marginalized: \\[4pt] profiled:} &  & \makecell[tr]{$0.0 \pm 1.6\times 10^{-2}$ \\[4pt] $0.0 \pm 7.0\times 10^{-2}$} & \makecell[tr]{$0.1 \pm 1.3\times 10^{-2}$ \\[4pt] $0.2^{+3.7}_{-3.9}\times 10^{-2}$} &  & \makecell[tr]{$0.0 \pm 1.6\times 10^{-2}$ \\[4pt] $0.0 \pm 7.0\times 10^{-2}$} & \makecell[tr]{$0.1 \pm 1.3\times 10^{-2}$ \\[4pt] $0.2^{+3.7}_{-3.9}\times 10^{-2}$} & \\
\end{longtable}

\end{document}